\PassOptionsToPackage{table,dvipsnames}{xcolor}
\PassOptionsToPackage{utf8}{inputenc}
\documentclass[sigconf, screen]{acmart}
\AtBeginDocument{%
  }

\setcopyright{acmlicensed}
\copyrightyear{2018}
\acmYear{2018}
\acmDOI{XXXXXXX.XXXXXXX}
\acmConference[xx]{xx}{June 03--05,
  2018}{Woodstock, NY}
\acmISBN{978-1-4503-XXXX-X/2018/06}
\usepackage[utf8]{inputenc}
\usepackage[T1]{fontenc}
\usepackage{xcolor}     
\usepackage{array}      
\usepackage{colortbl}   
\usepackage{booktabs}
\usepackage{graphicx}
\usepackage{subcaption}
\usepackage{float}
\usepackage{tabularray}
\UseTblrLibrary{booktabs}
\usepackage{longtable}
\usepackage{xcolor}
\usepackage{multirow}
\usepackage{enumitem}
\definecolor{mygreen}{RGB}{217, 234, 211} 
\definecolor{myyellow}{RGB}{255, 242, 204} 
\definecolor{myred}{RGB}{244, 204, 204} 

\newcommand{\redacted}[1]{\textbf{REDACTED}} 
\newcommand{\revision}[1]{#1}
\emergencystretch=\maxdimen
\newenvironment{s_itemize}{
\begin{itemize}[nosep,leftmargin=14pt]
}{\end{itemize}}
\newenvironment{s_enumerate}{
\begin{enumerate}[nosep,leftmargin=14pt]
}{\end{enumerate}}

\begin{document}

\title[How One-Minute Interventions Fit or Falter Across Domains]
{Designing for the Moment: How One-Minute Interventions Fit or Falter Across Domains}

\author{Zahra Hassanzadeh}
\email{hassanzadeh@cs.toronto.edu}
\orcid{0000-0002-5501-7972}
\affiliation{%
  \institution{University of Toronto}
  \country{Canada}
}
\author{Anne Hsu}
\email{anne.hsu@qmul.ac.uk}
\orcid{0000-0001-7987-9342}
\affiliation{%
  \institution{Queen Mary University of London}
  \country{United Kingdom} }
  
\author{Rachel Kornfield}
\email{rachel.kornfield@northwestern.edu}
\orcid{0000-0001-8542-6913}
\affiliation{%
  \institution{Northwestern University}
  \country{United States} }
  
\author{David Haag}
\email{david.haag@lbg.ac.at}
\orcid{0000-0002-9420-7111}
\affiliation{%
  \institution{Ludwig Boltzmann Institute for Digital Health and Prevention}
  \country{Austria} }
  
\author{Ananya Bhattacharjee}
\email{ananyabh@stanford.edu}
\orcid{0000-0002-9116-3766}
\affiliation{%
  \institution{Stanford University}
  \country{United States} }

\author{Jay Olson}
\email{jay.olson@mail.mcgill.ca}
\orcid{0000-0002-1161-5209}
\affiliation{%
  \institution{University of Toronto}
  \country{Canada} }
  
\author{Jan David Smeddinck}
\email{jan.smeddinck@dhp.lbg.ac.at}
\orcid{0000-0003-0562-8473}
\affiliation{%
  \institution{Ludwig Boltzmann Institute for Digital Health and Prevention}
  \country{Austria} }
  
\author{Norman Farb}
\email{norman.farb@utoronto.ca }
\orcid{0000-0002-8407-2938}
\affiliation{%
  \institution{University of Toronto}
  \country{Canada} }
  
\author{Alex Mariakakis}
\email{mariakakis@cs.toronto.edu}
\orcid{0000-0002-9986-3345}
\affiliation{%
  \institution{University of Toronto}
  \country {Canada}}

\author{Lydia Chilton}
\email{chilton@cs.columbia.edu}
\orcid{0000-0002-1737-1276}
\affiliation{%
  \institution{Columbia University}
  \country {United States} }

\author{Joseph Jay Williams}
\email{Williams@cs.toronto.edu}
\orcid{0000-0002-9122-5242}
\affiliation{%
  \institution{University of Toronto}
  \country{Canada} }

\renewcommand{\shortauthors}{xx}

\begin{abstract}
This paper explores the design space for one-minute digital interventions that prompt immediate action without onboarding or sensing. 
By embracing Fogg's Behavior Model and four design principles informed by literature, the goal of these interventions was to provide triggers that encourage actions so simple that even people with low motivation would be willing to complete them.
We examined the utility of these prompts by conducting a 14-day study with 22 participants interested in making small lifestyle improvements in at least one of three domains: physical activity, healthy eating, and mental well-being. 
Although we found evidence supporting the intent and design of our one-minute interventions, we also observed that participants encountered four types of barriers that hindered follow-through: temporal, physical, resource, and cognitive friction. 
When combined with insights drawn from participants' rewrites of our prompts, our findings suggest that intentional personalization through co-authorship could be a lightweight personalization mechanism that balances relevance with low friction.

\end{abstract}


\begin{teaserfigure}
\centering
    \includegraphics[width=\linewidth]{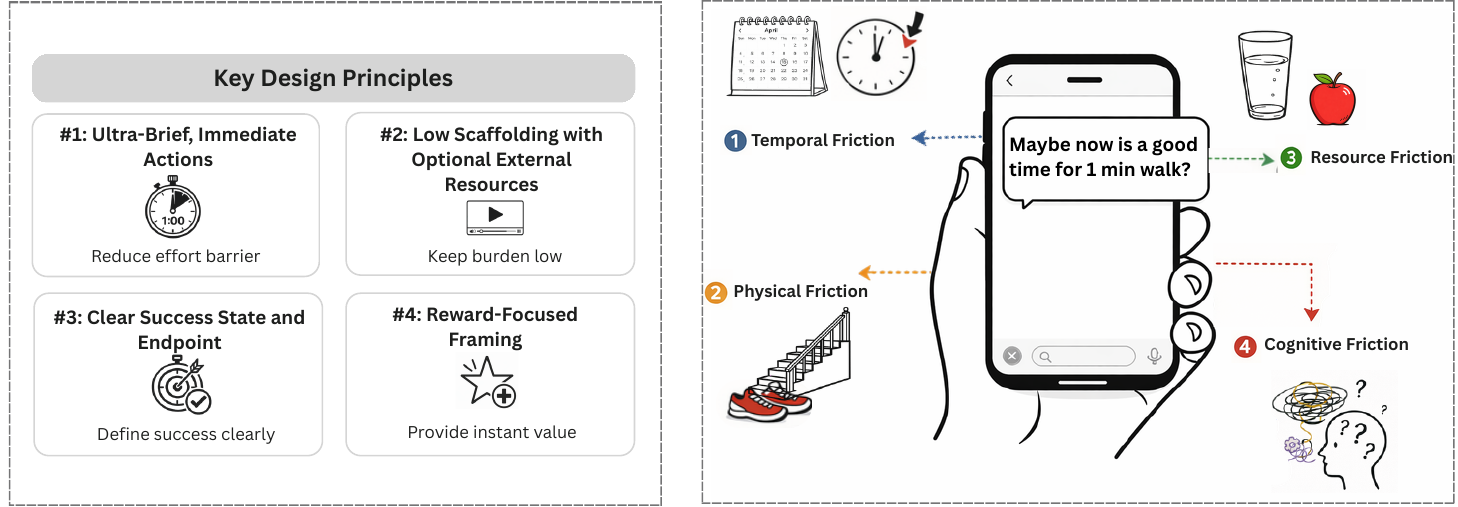} 
  \caption{(left) We grounded one-minute micro-interventions in four guiding design principles. (right) Our design probes revealed four categories of friction that shape whether ultra-brief interventions feel doable in the moment.}
  \Description{The left-hand illustration lists four key design principles for the intervention design: (1) Ultra-Brief, Immediate Actions to reduce effort barrier, (2) Low Scaffolding with Optional External Resources to keep burden low, (3) Clear Success State and Endpoint to define success clearly, and (4) Reward-Focused Framing to provide instant value.
  The right-hand illustration shows a smartphone held in one hand displaying a message that reads, ‘Maybe now is a good time for a 1-min walk?’ Surrounding the phone are four labeled categories of friction: (1) Temporal friction is represented by a calendar and clock, (2) physical friction by stairs and athletic shoes, (3) resource friction by a glass of water and an apple, and (4) cognitive friction by a head with question marks and tangled lines.}
  \label{fig:frictions}
\end{teaserfigure}

\maketitle
\section{Introduction} 
Mobile messaging platforms like SMS and WhatsApp have broad accessibility and low barriers to use, making them especially attractive channels for encouraging healthy behaviors like medication adherence \cite{thakkar2016text}, smoking cessation \cite{Liao2018}, and physical activity \cite{Head2013_textMessagesMeta}.
Such interventions have typically been envisioned as multi-week programs that build over time, requiring sustained engagement for users to appreciate the orchestrated experience.

This challenge has led some researchers to explore shorter alternatives such as single-session interventions~\cite{schleider2025ssi, bhattacharjee2025}, micro-interventions~\cite{Reza2023, dasilva2025, chau2025}, ecological momentary interventions (EMIs)~\cite{Heron2010, Balaskas2021EMIscoping}, and just-in-time adaptive interventions (JITAIs)~\cite{Nahum2017, Mishra2021}.
Still, many of these interventions require users to enroll in a program to generate a profile, and personalization is achieved either through extensive data collection or through algorithms that may encroach on people's sense of privacy. 
Such frictions reduce the reach and effectiveness of behavior change interventions across domains~\cite{Thaler2008Nudge}. 


Because of their high onboarding and commitment requirements, the aforementioned interventions have typically been designed for and evaluated with specific at-risk populations, such as sedentary individuals needing encouragement for physical activity~\cite{Luo2018, Li2025}, individuals with depression or anxiety symptoms~\cite{Schleider2017, kornfield2022}, or individuals with diabetes seeking guidance for healthy eating~\cite{Desai2019, mitchell2021, karimi2023}. 
Mobile messaging interventions may also be beneficial to the broader population, yet findings from prior research on user motivation and engagement may not generalize to everyday users. 
Individuals who do not perceive themselves as needing an intervention may be less likely to enroll in formal programs, thus limiting opportunities for onboarding, personalization, and progressive content delivery. 
This creates a gap in understanding how lightweight interventions can function without prior user data and support low-commitment engagement across multiple everyday health behaviors. 


Drawing on the Diffusion of Innovations Theory~\cite{Rogers2014}, our work prioritizes trialability—the ability to try an intervention with minimal effort, risk, or commitment. 
We investigate the design and utility of an extreme case of trialability through \textit{one-minute interventions} that prompt an action that can be completed in one minute without access to additional user information. 
By structuring prompts as standalone actions, one-minute interventions can be easily translated across behavior change domains, thereby allowing us to study \revision{how they fit or break down across domains}.
\revision{We do not treat one minute as a clinically established cutoff. Instead, we use it as a deliberately strict design constraint to study the lower bound of brief digital intervention design. If it were any shorter, the prompt might come across as a distraction rather than a meaningful intervention; if it were any longer, it might seem too burdensome to complete.}

This paper examines how one-minute interventions can align with everyday life across a range of behavior change domains. 
More specifically, we seek to answer the following research questions:
\begin{itemize}
    \item[\textbf{RQ1}] What are useful design principles for one-minute interventions? 
    \item[\textbf{RQ2}] How does the type of health behavior affect whether one-minute prompts work in practice?
\end{itemize}
We begin by explaining how recommendations from literature informed our initial set of design principles.
We then applied these principles to create one-minute interventions for three domains with significant relevance to both behavioral science and HCI: physical activity (e.g., stretching, marching in place)~\cite{Cambo2017, Luo2018}, healthy eating (e.g., snack planning, drinking more water)~\cite{zhang2021, Biel2018}, and mental well-being (e.g., mood check-in, gratitude reflection)~\cite{Caldeira2017, Diniz2023Gratitude}. 
While these domains are common targets of everyday behavior change efforts, \revision{they have divergent affordances for action,} allowing us to test whether the same design principles hold across distinct kinds of barriers and routines. \revision{This approach further enabled us to identify patterns that are universally generalizable across all domains as well as those that are domain-specific.}
Using the messages as design probes, we conducted a 14-day study to characterize the benefits and limitations of such interventions in terms of the kinds of actions being suggested and the circumstances under which they are received.
We also used the deployment to identify any underexplored benefits that such prompts may offer to general populations outside of formal programs.

Participants' \revision{reported prompt completion} during our study (82.5\% over 14 days) indicates that one-minute interventions are feasible to deploy and \revision{perceived as actionable in everyday life}. 
At the same time, feasibility did not mean frictionless use: even one-minute prompts could feel disruptive.
Our findings revealed that participants experienced four types of impediments that prevented action: temporal, physical, resource, and cognitive friction. 
The definition of trialability was therefore not strictly correlated with the hypothetical time required to complete actions.
Rather, it depended on domain-specific conditions: physical activity required physical context compatibility, eating hinged on narrow decision windows, and mental well-being demanded available cognitive capacity.
These results suggest that effective one-minute interventions still require selective personalization, but this adaptation can focus strictly on the variables that matter most for each behavior domain.
The main contributions of our research are:
\begin{s_enumerate}
    \item An exploration of the design space for one-minute interventions guided by our design principles: (1) ultra-brief immediate actions; (2) low scaffolding with optional external resources; (3) clear success state and endpoint; and (4) reward-focused framing.
    \revision{Our findings reveal frictions as initialization costs that can still make an ultra-brief prompt feel difficult to start, even when the action itself is simple.}
    \item An evaluation of our design principles using a 14-day probe deployment and semi-structured interviews \revision{across physical activity, healthy eating, and mental well-being. By comparing across domains, we show that one-minute prompts do not fail or succeed for one general reason: physical activity prompts were shaped more by space and bodily readiness, eating prompts by timing and available resources, and mental well-being prompts by cognitive effort and attention. We also show that allowing participants to rewrite prompts points to co-authorship as a lightweight form of personalization, where users can adapt generic prompts to their own tone, goals, and constraints.}
\end{s_enumerate}

\section{Related Work}
We first provide an overview of brief digital interventions for behavior change and health-promoting actions, after which we enumerate the persistent challenges that affect their uptake.
We then position our work relative to this literature.

\subsection{Digital Intervention Designs}
Multi-week behavior-change programs face persistent challenges with sustained engagement and attrition.
Despite their potential effectiveness, users frequently discontinue digital health interventions within days.
\citet{Baumel2019} observed median retention rates of only 3.9\% after 15 days across several mental health apps, while other studies report that users are twice as likely to drop out of digital interventions compared to traditional therapy~\cite{Waller2009}.
Common barriers include perceived time burden, lack of immediate relevance, and the overwhelming nature of long-term commitments.
This motivates interest in interventions that reduce the time and commitment required while maintaining effectiveness.

\subsubsection{Single-Session and Micro-Interventions}
Single-session interventions compress multi-session therapeutic content into one-time encounters lasting roughly 30–90 minutes. 
Core components typically include a clearly defined therapeutic target, brief psychoeducation grounded in an evidence-based principle, and a structured exercise wherein users apply the principle.
Typical exercises include but are not limited to reframing thoughts, generating advice, or planning adaptive actions to promote internalization and near-term change~\cite{schleider2025ssi}.

Micro-interventions are positioned as even shorter, modular sessions that can be revisited and sequenced within a broader behavior change programs~\cite{Baumel2020}. 
Frameworks like Design for Microintervention Software Technology describe architectures spanning an overarching narrative, a micro-intervention module (often 3–15 minutes), and constituent events or exercises~\cite{persson2025}. 
A growing body of evidence suggests that brief interventions can rival their longer counterparts in effectiveness, in part because simply getting started is often the hardest part of behavior change. 
Completing a small action through a brief intervention can reinforce self-efficacy by showing that change is achievable, and this self-efficacy can in turn strengthen behavior change~\cite{Bandura1997}.
These insights motivate exploring even briefer formats that minimize barriers to initial action.

\subsubsection{EMIs and JITAIs}
Building on the accessibility of messaging platforms, researchers have developed increasingly sophisticated delivery approaches. 
EMIs leverage messaging to provide real-world support as people go about their daily lives~\cite{Heron2010}, while JITAIs use contextual and sensor data to optimize the timing and content of messages~\cite{Nahum2017}. 
For example, systems like HeartSteps deliver personalized physical activity suggestions based on location, weather conditions, and prior activity patterns to maximize receptivity~\cite{liao2020}.

Sophisticated intervention systems like these come with substantial requirements that may limit their reach. 
JITAIs typically require sensing infrastructure, complex algorithms, and extensive user data collection~\cite{choi2019, kunzler2020, Mishra2021, Orzikulova2024} that may seem burdensome or invasive to prospective users.
Even simpler programs often require background information and sustained participation from users. 
While these features can enhance personalization and effectiveness, they also create barriers to initial adoption. 


\subsection{Factors Influencing Intervention Uptake}
Although many studies have shown positive outcomes from brief digital interventions, achieving these outcomes is not trivial and requires overcoming significant challenges in sustaining user engagement.

\subsubsection{Interventions as Interruptions}
From a human-computer interaction perspective, brief prompts delivered through messaging are delivered through an interruptive medium; they can capture attention and provide broad reach, but they also compete with ongoing activity and can create disruption or frustration if they arrive at the wrong time or demand too much coordination~\cite{CHOI2014}.
As a result, each prompt requires attention, interpretation, and a decision to act or defer while the user is engaged in whatever they are currently doing.
Conversely, interruptions can incur recovery costs during task resumption~\cite{Brumby2013}. 
Prior work on interruption management shows that aligning interruptions with task breakpoints can reduce perceived disruption and improve user experience~\cite{IqbalBailey2008}. 
Field studies further show that users often use alerts for passive awareness and may defer action rather than immediately switching tasks~\cite{IqbalHorvitz2010}, highlighting the challenge of designing prompts that are worth acting on as soon as they arrive. 

\subsubsection{Personalization}
Interventions that rely on generic or impersonal messages may fail to resonate with users' evolving needs and situational constraints \cite{Bhattacharjee2022-2, Slovak2023}. 
Furthermore, digital health tools compete for attention in saturated digital environments, where even personalized content can be ignored if not delivered at the right moment \cite{folstad2018different, Muench2017}. 
Recent work in human-computer interaction and health informatics emphasizes the importance of personalization, timing, and contextual awareness in message design. 
For example, systems such as \emph{Snack Talk} have shown that embedding behavioral prompts into familiar family routines can promote sustained engagement by leveraging social support and natural context \cite{Bateman2022SnackTalk}. 
Related work on adaptive delivery via JITAI further reinforces the value of aligning prompts with users’ momentary circumstances.
Although formative investigations into scalable and deep personalization based on generative AI are emerging \cite{Haag2025}, JITAIs remain difficult to deploy. 



\subsection{Our Goals}
Beyond clinical or condition-specific programs, prior work has demonstrated promise for messaging-based support of everyday health-promoting behaviors. 
For example, healthy-eating interventions such as \emph{Intuitive Eating SMS}~\cite{Manana2023} and \emph{Txt4HappyKids}~\cite{Power2017} use personalized messages to promote mindful eating and nutrition guidance, with positive user perceptions and preliminary evidence of behavior change.
Interventions to reduce screen time similarly use prompts to encourage healthier digital habits, including school-based and family-centered approaches that demonstrate feasibility and acceptability~\cite{wahi2011screen, tomayko2021}.
In parallel, health science research shows that short, repeatable actions (e.g., “exercise snacks” or vigorous intermittent lifestyle physical activity) can provide meaningful physiological benefits~\cite{rooksby2014personal, jacobson2001}, suggesting that brief messaging prompts could encourage beneficial micro-behaviors throughout the day.

Many everyday health behaviors include both planned and habitual components~\cite{wood2016, pinder2018}. 
For example, drinking water throughout the day can become a habitual response to certain contexts (e.g., seeing a water bottle, finishing a meal), while choosing to eat a healthy snack requires a deliberate decision at each opportunity.
Habit theory helps explain why context, cues, and environmental stability can strongly shape whether a person acts, even when motivation is high~\cite{wood2016, verplankenwood2006}. 
While habit formation is an important long-term goal for sustained behavior change, our study focuses on in-the-moment enactment.
We make this choice for two reasons.
First, understanding immediate uptake is a prerequisite to studying habit formation; if people cannot or will not act on a prompt in the moment, there is no opportunity for repetition to build habits.
Second, users may benefit from occasional prompted actions even without forming stable habits, making in-the-moment enactment valuable in its own right.

The literature leaves important questions open about the design of very brief interventions as a unit of interaction, especially when they must work without sensing infrastructure or intensive personalization.
We still have limited guidance on which design elements matter most \revision{once the target action is confined to a single minute}.
It also remains unclear how these design choices interact with real-world barriers that may differ by behavior domain.
Finally, much prior work evaluates bundled programs or delivery policies, making it difficult to isolate which prompt-level design decisions shape in-the-moment uptake.
Our work addresses this gap by empirically studying one-minute messaging prompts as a highly constrained form of micro-intervention.
We seek to understand both the limitations of extreme brevity and any underexplored benefits in everyday contexts.

\section{Guiding Design \revision{Objectives}}
Our work pushes the limits of what is known about brief digital interventions by pushing their duration to one minute and eliminating the need for intervention-specific infrastructure. 
We used the following principles established in previous work to inform our prompt designs.

\subsection{\revision{Objective} \#1: Ultra-Brief, Immediate Actions} 
Prior work on micro-interventions has noted that people can struggle to initiate activities when they feel too large, require preparation, or must be scheduled \cite{Nahum2017, RennickEgglestone2016}. 
The Fogg Behavior Model \cite{fogg2009} further highlights that behavior occurs when receipt of a prompt coincides with sufficient motivation and ability to carry it out, suggesting that making the required action manageable in scope and immediate is one way to increase ability at the moment of the prompt. 

\smallskip
\noindent\textbf{Design \revision{Objective}:} One-minute interventions should be ultra-brief and obviously executable in the moment without scheduling or preparation.

\smallskip
\noindent\textbf{Initial Implementation:} We limited each intervention to a single action that takes about 60 seconds and is framed as something that can be done immediately.
For example:
\begin{quote}
\texttt{“Roll your shoulders, stretch your neck, and release tension—right now. Movement relieves stress and refreshes your focus. Follow this quick 60-second stretch routine.”}
\end{quote}

\noindent
\revision{
The one-minute duration refers to the target action rather than the full interaction time that includes reading the prompt and potentially typing a response, although those steps were expected to be equally brief.
This design objective was intended to study a lower-bound case of brief intervention design: an action short enough that it should not require significant preparation or commitment, yet long enough to have a visible start and endpoint. 
This boundary also helped us examine whether reducing action duration is enough to make a prompt feel doable in daily life. 
If a one-minute action still failed, we could look beyond duration and identify other forms of burden. 
}

Across all domains, we avoided language that suggested planning, tracking, or scheduling. 
With \textit{immediate-action prompting}, we used direct language to convey that the action could be started immediately and completed in the next minute to support task initiation.

\subsection{\revision{Objective} \#2: Low Scaffolding with Optional External Resources}
Micro-interventions and JITAIs often require substantial scaffolding (e.g., apps, dashboards, progress graphs, and rich feedback) to support engagement and reflection over time~\cite{Nahum2017, MilneIves2020}. 
These features can be helpful, but they also add time and cognitive load to each interaction and assume infrastructure that is not always available. 

\smallskip
\noindent\textbf{Design \revision{Objective}:} One-minute interventions should require minimal scaffolding by avoiding dedicated app installation or multi-step programs. Additional resources should be optional to support maintenance of the one-minute action when needed.

\smallskip
\noindent\textbf{Initial Implementation:} 
We deployed our prompts via WhatsApp, a platform used by billions worldwide with features comparable to those of most other basic messaging services.
Messages were randomly assigned to users and sent independently such that there was no need to track dependencies.

In terms of optional resources, our prompts often included a link to a 60-second video of a person providing one of two forms of scaffolding in an enthusiastic tone.
The first was support for context switching, such as:
\begin{quote}
\texttt{“Why not try closing your eyes for 60 seconds? Just close them now to block out any stimulation and let your mind wander. Click on this video and follow along [Video Link]. It might not feel magical, but it’s a small, helpful way to support your mind and body.”}
\end{quote}
The second was instructional scaffolding that provided users with explicit guidance on how to complete the action, for instance:
\begin{quote}
\texttt{“How about taking 1 minute to stretch or do a short exercise? Show your arms, neck, and legs some love. Just click this link for a quick exercise video and follow along: [Video Link]”}
\end{quote}
Regardless of the scaffolding provided, the message text was intended to be sufficient on its own, meaning that users could complete the intervention without opening the video. 

\subsection{\revision{Objective} \#3: Clear Success State and Endpoint} 
Action-phase models emphasize that there is often a gap between starting a task and knowing when it is finished \cite{HeckhausenGollwitzer1987, Gollwitzer1990}. 
Ambiguous goals like \texttt{“be more mindful”} and \texttt{“move more”} make initiation and completion hard to recognize, which weakens the sense of competence and makes repetition less likely. 

\smallskip
\noindent\textbf{Design \revision{Objective}:} One-minute interventions should specify a clear success state so that users know exactly when the task starts and when it is complete.

\smallskip
\noindent\textbf{Initial Implementation:} We defined success in observable, time-bounded terms. Text- and video-based instructions were aligned so that completion of the video segment corresponded to completion of the task. 
For example:
\begin{quote}
\texttt{“Take a quick break and click on this link [Video Link]. Just let the video guide you while you put your screen aside. Trust me, it’s a small break that might make a big difference.”}
\end{quote}
Users were also encouraged to indicate completion by replying with a brief message (e.g., \texttt{“Done”}, \texttt{“Complete”}), providing a low-effort way of demonstrating accomplishment.

\subsection{\revision{Objective} \#4: Reward-Focused Framing}
The Fogg Behavior Model~\cite{fogg2009} highlights that behavior occurs when a prompt coincides with sufficient motivation and ability. 
While \revision{Objective} \#1 addresses ability by making actions immediately executable, this \revision{objective} addresses motivation. 
Moreover, literature on reward circuits shows that value, attention, motivation, and action are tightly coupled \cite{BissonetteRoesch2016}, as directing people's attention to specific goals changes how attractive options feel in the moment.
For example, nudging people to focus on health increases the importance of health attributes in their choices \cite{Hare2011}. 
By explicitly highlighting the benefits of an action (e.g., "movement relieves stress"), we aim to increase momentary motivation when the prompt arrives.

\smallskip
\noindent\textbf{Design \revision{Objective}:}
Before or while the action is suggested, one-minute interventions should direct attention to the reward of doing it, thereby increasing momentary motivation and briefly tilting the user’s valuation system toward the beneficial outcome. 

\smallskip
\noindent\textbf{Initial Implementation:}
Beyond using encouraging and explicit language to highlight the benefits of taking action, we explored the idea of \textit{reflection-first prompts} to draw people's attention to potential rewards.
Borrowing from the spirit of motivational interviewing \cite{MillerRollnick2013}, these prompts asked users to consider their current state and whether it could be improved within the next minute, evoking why the action might matter to them in the moment. For example: 
\begin{quote}
\texttt{“What emotion is sitting with you right now? Take a deep breath and notice, are you feeling calm, stressed, or something in between? Watch this 60-second grounding video to help bring clarity to your emotions. Small check-ins like this strengthen emotional awareness and reduce mental fatigue.[Video Link].”}
\end{quote}

\section{Study Design}
We conducted a 14-day study to identify when and why seemingly simple one-minute interventions might succeed or break down in daily life. 
This protocol received approval from the Research Ethics Board at the University of Toronto. 


\subsection{Participants}
We recruited participants through Prolific to gather feedback from diverse individuals with varied life experiences and routines.
The call for participation invited adults who are interested in receiving encouragement to better their lifestyle with respect to our three target domains: physical activity, healthy eating, and mental well-being. 
Participants were also compensated \$20 USD for enrolling in the study.

We enrolled 22 adults (13~women and 9~men) living in North America.
Participants' ages ranged from 18 to 64 years according to the following distribution: 18-24 (n~=~1), 25–34 (n~=~9), 35–44 (n~=~6), 45–54 (n~=~4), and 55-64 (n~=~2).
Ten participants were employed full-time, 5 were employed part-time, 3 were unemployed and seeking work, 3 were unemployed and not seeking work, and 1 was retired. 
Their educational backgrounds ranged from high school diplomas to advanced degrees.

\subsection{Procedure}
Participants were randomly assigned to one of two cohorts in a within-subjects study design.
Half of the participants received immediate-action prompts during the first week and reflection-first prompts during the second week, while the other participants experienced the reverse order.
\revision{Since all participants experienced both prompt types, counterbalancing helped avoid confounding prompt type with first-week novelty, study fatigue, or learning.}
The full list of prompts we created is provided in Appendix~\ref{appendix:study2-prompts}.

\revision{Each participant received one prompt per day for 14 days via WhatsApp. Prompts were delivered in one of three broad time windows that roughly aligned with typical daily rhythms — 10 AM–12 PM for the morning, 3–6 PM for the afternoon, and 7–10 PM for the evening — using a chatbot that followed the procedural flow illustrated in Appendix~\ref{appendix:flow}.
We rotated the delivery window across days using a pre-specified schedule so participants received prompts at various times during the study period.
One-minute interventions were assigned using permuted blocks to ensure balanced exposure to all three target domains while maintaining randomization within each block. Each prompt included a short text message describing a concrete one-minute action, along with a link to an optional companion video.}

\subsection{Data Collection}
Whenever participants completed a task, they were encouraged to type "Done" into the WhatsApp chat to indicate completion.
\revision{Completion replies were self-reported acknowledgments; we did not conduct any independent verification to confirm whether the behavior had happened, although there were no indicators of dishonest self-reporting during our interactions with participants.}
Participants received a reminder if they did not respond to a prompt within 2 hours of receiving the message.
Participants were also sent invitations on select days (Days 3, 5, 9, and 14) to reply with suggested revisions to the tone and content of the message they had just received.
These suggestions were not incorporated into future messages, but instead captured people's in-the-moment reactions to receiving the prompt.
The amount of compensation participants received was not contingent on their response rate to any of these requests.

Within one week after the study, participants were invited to complete a semi-structured interview via Google Meet; 12 (54.5\%) accepted the invitation.
Each session lasted approximately 30–45 minutes.
The goal of the interview was to understand participants' subjective experiences with different types of one-minute interventions, the messages that were most or least impactful, and their thoughts on how message design influenced their engagement. 
Examples of interview questions included the following:
\begin{s_itemize}
    \item What was the most helpful or memorable prompt you received during the 14-day period? Why did it stand out to you?
    \item Were there any moments when you felt less motivated to follow the prompt? Can you describe what was going on at the time?
    \item Did you find any difference between the types of messages? How did they feel different?
    \item Can you recall any instance when the prompt led you to reflect more deeply on your health or well-being?
    \item Were there any specific days or message types that felt harder to engage with? Why?
\end{s_itemize}

\subsection{Analysis}
After transcribing the interviews, we applied thematic analysis~\cite{Braun2019} to both participants' message responses during the study and the interview transcripts afterwards.
Our process was guided by our research questions and design principles, with pre-existing codes based on the core elements of actionability, engagement, and perceived benefits.
To strengthen credibility, a second researcher independently reviewed roughly 20\% of the transcripts before questioning assumptions and refining surfaced themes.
Participant quotes and opinions are attributed with the identifier P\#.

We also quantified participants' engagement with the prompts according to their average video click-through rates and task completion rates.
These numbers are not intended to provide a definitive measurement of engagement, but rather to assure us that the participants were sufficiently engaged to provide feedback on the one-minute interventions.

\section{Findings}
In this section, we first provide a brief overview of how often participants engaged with the one-minute interventions throughout the study and of their positive experiences.
We then address two major factors that shaped participants' perspectives of the interventions: the cost of interrupting one's current activity and the perceived relevance of the suggested activity.

\subsection{Engagement}
\begin{table}[!tp]
 \caption{A summary of prompt engagement metrics during our 14-day study.}
  \label{tab:engagement-metrics}
  \centering
  \begin{tabular}{@{}lcc@{}}
    \toprule
    \textbf{Target Domain} & 
    \textbf{\begin{tabular}[c]{@{}c@{}}Task Completion\\ Rate\end{tabular}} & 
    \textbf{\begin{tabular}[c]{@{}c@{}}Video Click-Through\\Rate\end{tabular}} \\
    \midrule
    Mental Well-being      & 86.6\% & 80.0\% \\
    Physical Activity      & 90.5\% & 79.7\% \\
    Eating Healthy         & 87.1\% & 70.0\% \\
    All                    & 87.2\% & 76.9\% \\
    \bottomrule
  \end{tabular}
\end{table}

\begin{figure}[t]
  \centering
  \includegraphics[width=\linewidth]{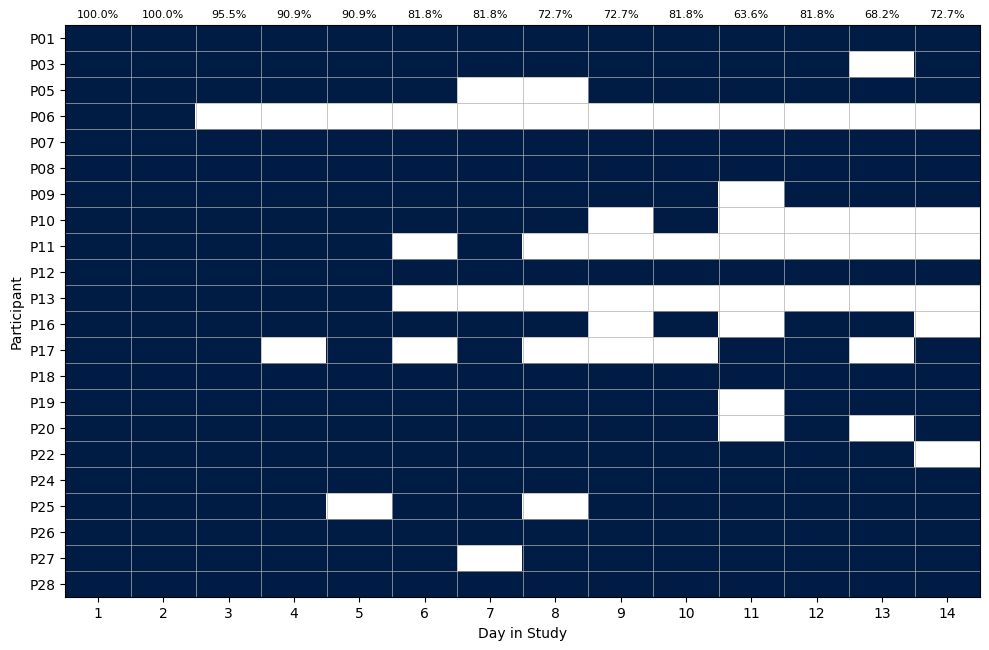}
  \caption{A heatmap showing how participants engaged with the one-minute interventions over the 14-day study period. Blue cells indicate days when participants \revision{self-}reported completing an intervention, while empty cells indicate days without a response.}
  \Description{A grid where each row represents one of the 22 participants who completed the study, and each column represents one of the 14 days in the study. Each cell is filled whenever a participant self-reported completing a prompt for the given day.
  Engagement was high early, but then waned for some participants later in the study.}
  \label{fig:heatmap}
\end{figure}

\autoref{fig:heatmap} illustrates how often participants \revision{self}-reported completing the one-minute interventions throughout the study, while \autoref{tab:engagement-metrics} reports the average engagement metrics disaggregated by target domain.
Across all prompts, participants clicked on 223 (72.4\%) of the video links and reported completing 254 (82.5\%) of the tasks. 
Note that the completion rate exceeded the video click-through rate because participants could, and often did, complete the suggested action without opening the video and instead relying on the text description.
Among the 22 participants, 8 participants demonstrated consistent adherence by responding to the prompts on all 14 days. 
Two participants disengaged entirely within the first 7 days, while two participants dropped off in the second week.

Our key takeaway from these observations was that participants were sufficiently engaged to provide design feedback. 
While we hypothesize that engagement was high
due to the low-friction nature of the interventions, we also acknowledge that the numbers may be inflated by the Hawthorne effect \cite{McCarney2007}.
\revision{We also cannot claim that behavior change occurred due to the limited scope of our study.}
Therefore, we hesitate to make strong claims based on these numbers and instead turn to our qualitative data for more reliable insights.

\subsection{Positive Experiences}
Participants appreciated several aspects of the one-minute interventions.
Successful prompts were often characterized as clear, simple, and easy to integrate into ongoing routines. 
Participants referred to the activities as \textit{“easy to follow”} (P7), \textit{“simple tasks that made me feel good and broke up the monotony in my day”} (P10), and \textit{“not requiring a lot of time, but something you can do easily”} (P9). 
They also noted that the bounded duration and clarity of the prompts made them manageable even on busy or low-energy days.

Participants who engaged with the prompts reported positive emotional and physical outcomes after completing tasks.  
For example, P1 stated, \textit{“I felt great! I had more energy to get up and go shopping.”}
These feelings were attributed not only to the perceived short-term emotional and physical benefits of the interventions but also to the satisfaction of completing a small task.
Reflecting on the WhatsApp notification's looming presence as a reminder of an unfinished task, P5 noted, \textit{“The icon kept staring at me, so I eventually did it”}.

An unexpected finding was that when a prompt did not fit their immediate mood or situation, some participants adapted it into an alternative one-minute action rather than skipping it entirely. 
As P25 explained, \textit{“I didn’t have the mental space for the one about breathing, but I still drank water.”} 
This suggests that the prompts still provided some benefits even when participants were unable or unwilling to complete the suggested action.

\begin{table*}[!tp]
 \caption{A summary of the frictions participants described after receiving one-minute prompts across domains.}
  \label{tab:friction}
  \centering
  \footnotesize
  \setlength{\tabcolsep}{3pt}
  \renewcommand{\arraystretch}{1.5}
  \begin{tabular}{@{}p{0.08\textwidth} p{0.29\textwidth} p{0.29\textwidth} p{0.29\textwidth}@{}}
    \toprule
    \textbf{Frictions} & 
    \textbf{Physical Activity} & 
    \textbf{Healthy Eating} &
    \textbf{Mental Wellbeing} \\
    \midrule
    Temporal & Timing often conflicted with work, classes, commutes, or time spent outside, making even brief exercises inconvenient at the time of delivery (P3, P9, P17, P18, P19, P20, P22, P24, P25, P26, P28). 
    & When prompts were received at moments distant from snacks and mealtimes, participants reported they were not actionable and required extra planning (P17, P22). 
    & Prompts were sometimes received during work or busy periods, when participants felt they could not pause despite finding the prompt useful (P8).\\
    Physical & Some prompts required stairs, space, or freedom of movement that participants lacked in their current environment (P19, P22, P25). 
    & No clear evidence 
    & Some participants only wanted to complete these prompts in a quiet or private space, which was not always accessible (P25). \\
    Resource & No clear evidence 
    & Health food and water were not always immediately available, so some prompts required grocery planning or prepared ingredients (P17, P22, P25). 
    & Some prompts required additional resources like headphones to listen to instructions or block ambient sound (P25). \\
    Cognitive & Exercises required a mental shift from sedentary or task-focused states, which participants found harder than expected for a one-minute activity (P26).
    & Some prompts were perceived as long-term behavior changes rather than quick actions, making them feel cognitively heavier than intended (P17).
    & Reflection-first prompts were frequently described as mentally demanding, extending beyond the expected one-minute scope and encouraging postponement (P25, P26, P9, P24). \\
    \bottomrule
  \end{tabular}
\end{table*}

\subsection{Importance of Interruption Costs}
Even when suggested activities were perceived as brief, participants deferred or skipped them when the cost of interrupting their current activity felt too high.
Although prior work has already reported these challenges~\cite{Webb2006, Kwasnicka2016, Feil2023}, our data analysis revealed four recurring frictions: temporal, physical, resource, and cognitive. 
\autoref{tab:friction} summarizes how these frictions manifested across the three domains, and we define them below:
\begin{s_enumerate}
\item \textbf{Temporal friction} reflected poor timing relative to participants' current activities and constraints.
Activities were easily dropped when they conflicted with activities like meetings, commutes, or bedtime.
\item \textbf{Physical friction} emerged when the suggested action mismatched participants' current physical context or ability.
Actions requiring a different environment caused minor inconveniences sufficient to deter engagement.
\item \textbf{Resource friction} occurred when prompts assumed preparation or access to equipment. 
Even small requirements, such as having a healthy snack or water nearby, introduced dependencies that made the action feel less immediate.
\item \textbf{Cognitive friction} appeared when participants perceived that the mental effort required for a prompt exceeded one minute. 
This was particularly true for reflection-first prompts, as P25 noted, \textit{“Reflection required more cognitive effort.”}
Some participants suggested that these prompts extended beyond their stated scope, making them feel that the task required more effort.
\end{s_enumerate}

When faced with these barriers, the suggested activities were at least simple enough for participants to tell themselves that they could defer them to a later time.
However, most participants acknowledged that they rarely revisited the prompts.
Our preprogrammed two-hour reminders were helpful to some, but were so easy to ignore that most never revisited the prompts.
For instance, P8 received a prompt to complete breathing exercises when working at a medical clinic. 
Although they recognized the value of the nudge at a time of day when they needed it most, they were so busy with their workload that they felt that they never had the opportunity to follow through. 
Below, we highlight how these frictions manifested across our target domains.

\subsubsection{Physical Activity}
Many participants were willing to incorporate physical activity at any time of day, yet what stopped them was physical frictions imposed by their surroundings.
Participants emphasized that universally accessible activities like posture checks or short walks succeeded because they could be done in the moment, while failures clustered around actions that required moving to a larger space or finding stairs. 
For example, P28 noted, \textit{“I couldn’t do the stretching one — switching from thought to body is hard.”}

\subsubsection{Healthy Eating}
Even though the one-minute interventions related to healthy eating were geared towards light snacks and water, the prompts were most successful when aligned with food-related decision-making before meals. 
When this was not the case, immediate-action prompts that emphasized simplicity were viewed as being more laborious than intended because they required added planning, which P25 claimed to be \textit{“not doable right away, [so you] may forget.”} 
In contrast, reflection-first prompts were more successful in this domain because they explicitly suggested brief preparation before taking action. 


\subsubsection{Mental Well-Being}
Participants described short breathing or grounding exercises as feasible in the moment, but there were conflicting views on the cognitive effort required to engage in them.
On the one hand, the instructions in these prompts were often considered helpful in reducing cognitive load by providing guidance that allowed participants to concentrate on their thoughts. 
On the other hand, some participants extended the activities into longer thought exercises that made future prompts seem less approachable or convenient. 
These prompts were also more difficult for participants to complete if they required quiet time or focused attention.

\subsection{Importance of Perceived Relevance}
Regardless of whether there was an explicit suggestion to reflect on the prompt, many participants considered whether the suggested activity addressed a current struggle before deciding whether to follow through. 
The same prompt was often viewed as highly motivating or completely irrelevant depending on whether it targeted an active struggle or an already-optimized behavior.
For example, P24 said that prompts to drink water were not very useful because they had been vigilant about doing so earlier in the day.
Meanwhile, reminders to maintain good posture resonated with individuals actively experiencing back or neck discomfort.

When participants were invited to rewrite selected prompts, their edits revealed efforts to shape the content and tone of the support they desired. 
Their revisions typically fell under one of three categories. 

\subsubsection{Tone and Emotional Fit}
Participants favored prompts that felt supportive and low-effort, particularly those tailored to physical activity or mental well-being.
Instead of standalone instructions, they preferred guided, conversational delivery that reduced cognitive demand.
For example, P19 expressed a desire for \textit{“slowing it down and making it more of a guide instead of an instruction~\dots Something you could follow along with instead of instructions you need to remember.”}
Rewrites in this category often added a friendly or inclusive tone (e.g., \texttt{“Hey y’all! How ya feeling?”} by P8) or drew from participants' ambient environment (e.g., \texttt{“A video filmed outdoors might be nice, with birds chirping”} by P28). 

\subsubsection{Personal Relevance and Motivation}
Participants modified prompts to emphasize tangible or emotionally salient short-term benefits, reflecting an effort to connect the task with something that felt personally helpful or needed in the moment.
This sentiment was best conveyed by P3, who stated, \textit{“It would be more helpful if it reminded me of how I’m feeling or what I need today.”}
Rewrites in this category typically acknowledged participants' recent actions and connected suggestions to personal goals (e.g. \texttt{"I know you drink water a few times a day, but are you actually hydrated?" by P8}).


\subsubsection{Identity Work}
A few participants' revisions went beyond adding personal relevance, using the one-minute interventions as an opportunity for identity expression. 
For example, P8 rewrote a healthy eating message into a personal challenge to try alternative beverages: \texttt{"Is anyone feeling more lethargic or even a ‘crash’ after sugary sodas or energy drinks? I challenge you to try these nutritious and energizing beverages for 30 days, and tell me how you feel then."}
In another instance, P28 said the following when linking a suggested gratitude exercise to a personal aspiration: \textit{"I wrote down a short list of what I’m currently grateful for. I would love to be a more grateful person."} 
Even if participants did not follow through on the suggested activities, the prompts provided an opportunity to reflect on the kinds of individuals they aspired to be. 
\section{Discussion}
In this section, we revisit our research questions by analyzing how our design principles performed in practice, examining cross-domain friction patterns that emerged, and discussing their implication for one-minute intervention design. 
We then discuss the limitations of our methods and opportunities for future investigation.

\subsection{RQ1: Revised Design Principles for One-Minute Interventions}
We informed the initial design of our one-minute interventions by drawing inspiration from human-computer interaction and psychology literature.
One of the biggest influences was the Fogg Behavior Model~\cite{fogg2009}, which states that prompts are more likely to succeed when there is sufficient motivation and ability to perform the target behavior.

To increase ability, all of the prompts described actions that could be completed in one minute.
Participants largely agreed that the suggested activities were simple, and the lightweight scaffolding provided by the video links was not overbearing.
However, research on digital notifications has shown that disruptions have a cost. 
People delay, forget, or fail to resume when the task does not fit the moment, even if it is small~\cite{IqbalHorvitz2010}. 
Our work extends this literature by identifying several frictions that made the one-minute interventions seem like disruptions despite their simplicity.
Physical, resource, and cognitive frictions were viewed as initialization costs that extended the perceived scope of suggestions beyond a minute (e.g., needing space, materials, or mental effort before starting).
Meanwhile, temporal frictions served competing demands that typically overshadowed the one-minute interventions because of their perceived importance.

To increase motivation without requiring significant infrastructure, we experimented with reflection-first prompts that encouraged people to first consider the benefits of taking action.
While this language enhanced the perceived relevance of messages, it also induced cognitive friction that occasionally deterred engagement.
Motivation was also influenced by how often participants had already engaged in the suggested activity.
When participants had recently completed the activity or were in the middle of doing it, the messages were viewed as irrelevant and unnecessary.
When it had been a while since they completed the activity, the interventions prompted action in the best-case scenario and served as reminders otherwise.

\revision{Our findings suggest that one-minute prompts are most appropriate when they make very few assumptions about the user's current activity, physical location, or the availability of necessary materials. 
They worked better when the action was familiar and easy to initiate, such as neck stretches, posture checks, breathing exercises, or drinking water. 
Participants described these prompts as accessible, time-efficient, and useful for breaking up the day. 
In contrast, prompts were more challenging to complete when they implicitly assumed too much, such as access to stairs, adequate physical space, healthy food nearby, or sufficient mental energy for reflection. 
This suggests that one-minute interventions should be designed to be robust across diverse ordinary situations: low-risk, low-preparation, flexible, and clear enough that users can either execute the action as written or slightly adapt it without compromising the intended benefit.
}

Putting these takeaways together, we revisit our guiding design principles to emphasize not only brevity but also fit with everyday constraints. 
The extended principles below focus on prompt-level design choices, with a minimal role for context captured through lightweight user control.

\subsubsection{Principle \#1: Ultra-Brief, Immediate Actions:} One-minute interventions should be ultra-brief and obviously executable in the moment without scheduling or preparation. 

\smallskip
\noindent\textbf{Extension:} While temporal brevity remains important, our findings demonstrate that brevity alone does not eliminate interruption costs.  
Physical friction emerged when prompts assumed environmental affordances participants lacked (e.g., stairs, open space), consistent with research showing that context mismatch reduces engagement~\cite{choi2019,kunzler2020}. 
Resource friction appeared when prompts assumed access to materials without verification (e.g., healthy snacks). 
Cognitive friction manifested when mental effort exceeded expectations, particularly for reflection-first prompts in the mental well-being domain. 
These findings suggest that while brevity addresses one dimension of ability in the Fogg Behavior Model~\cite{fogg2009}, domain-specific initialization costs represent a distinct barrier requiring separate design consideration.
Effective ultra-brief interventions must therefore address not only duration but also the conditions required to begin.

\subsubsection{Principle \#2: Low Scaffolding with Optional External Resources:} One-minute interventions should require minimal scaffolding by avoiding dedicated app installation or multi-step programs. Additional resources should be optional to support maintenance of the one-minute action when needed.

\smallskip
\noindent\textbf{Extension:} Our findings showed that optional video resources were frequently used (72.4\% click-through rate overall) and helpful. However, clicking and watching created resource friction and deferral points where participants delayed action. This highlights a tension between providing supportive resources and maintaining the immediacy that makes one-minute interventions actionable.
Designers should ensure messages are fully sufficient for task completion so videos remain truly optional; videos should enhance rather than enable the action.

\subsubsection{Principle \#3: Clear Success State and Endpoint:} One-minute interventions should specify a clear success state so that users know exactly when the task starts and when it is complete. 

\smallskip
\noindent\textbf{Extension:} Time-bounded instructions worked well for immediate-action prompts, and video durations aligned with task completion provided clear endpoints. 
However, reflection-first prompts sometimes exceeded one minute when participants engaged in multi-step reasoning, particularly regarding mental well-being. 
To maintain clear boundaries, designers should either keep reflections minimal or explicitly offer the option for extended suggestions rather than embedding complexity within the stated one-minute frame.

\subsubsection{Principle \#4: Reward-Focused Framing:} Before or while the action is suggested, one-minute interventions should direct attention to the reward of doing it, thereby increasing momentary motivation and briefly tilting the user's valuation system toward the beneficial outcome. 

\smallskip
\noindent\textbf{Extension:} Reward-focused framing through reflection-first prompts creates trade-offs. 
While reflection enhanced perceived relevance and personal connection, it also induced cognitive friction that occasionally affected engagement. 
This trade-off varied by domain: reflection-first prompts worked well for eating, where planning was helpful, but poorly for mental well-being, where reflection expanded cognitive load beyond one minute. 
Designers should balance reflection with cognitive feasibility for their target domain and consider supporting lightweight user feedback to avoid redundant prompts that reduce perceived autonomy.

\revision{One factor that became clearer from the study is that participants not only asked themselves if they could do the prompt; they also asked themselves whether it was wor
th doing it right away. A prompt could be short but still incur high perceived costs if it interrupted work, required physical displacement, necessitated specific resources, or demanded reflection during periods of cognitive depletion.
At the same time, participants were more willing to act when the benefit was evident and relevant. 
Therefore, shifting the cost-benefit trade-off in favor of a one-minute intervention depends not only on reducing temporal demands but also on making the action's immediate rewards perceptible at the moment of delivery. The four frictions we identified can be understood as the costs of this trade-off, while immediate relief or personal relevance form the benefits.}

\subsection{RQ2: Domain-Specific Considerations}
Unlike previous work, which is often situated in a single domain, we examined three domains simultaneously to elicit both unique and generalizable design considerations.
Doing so revealed that while timing and context are consistently crucial for effective interventions across all domains, the specific types of contextual information needed and the complexity of intervention design may vary significantly depending on the behavior being targeted.

\subsubsection{Physical Activity}
Physical activity prompts were often perceived as more immediately doable than prompts in other domains but their success still depended on timing and, especially, on whether the current physical context supported movement. 
The primary barrier was physical context: prompts that assumed a particular environment (stairs, open space, freedom of movement) failed when participants could not meet those conditions. 
Cognitive friction also appeared occasionally, as switching from a sedentary or task-focused state to physical movement required a mental shift that participants found harder than expected. 
This domain may be the most forgiving in terms of personalization requirements. 
Prompts can remain simple, but should offer ability-compatible alternatives within the prompt itself (e.g., \texttt{"Can't do stairs? Walk in place"}) so participants can select an action that fits their immediate context.

\subsubsection{Eating}
Healthy eating prompts showed the strongest dependence on two personalization variables: current baseline behavior and hitting narrow decision windows. 
The prompts became redundant when someone was already eating or drinking, and immediate-action prompts were only successful at specific moments of the day (e.g., planning, shopping, ordering).
Reflection-first prompts were more promising in this domain precisely because they encouraged brief planning before action, which aligned better with how many people plan their meals. 
Drawing on implementation intention research showing that advance planning increases follow-through~\cite{GOLLWITZER2006}, our findings suggest that eating interventions require a more complex design response.
Example modifications include lightweight behavioral profiling to avoid redundancy, time-of-day targeting to hit decision windows, and staged delivery (e.g., a message encouraging planning in the morning, followed by a message encouraging action closer to mealtime) to structure the intervention around the natural rhythm of food behavior.

\subsubsection{Mental Well-Being}
Mental well-being prompts demonstrated heightened sensitivity to users' transient emotional states and cognitive burdens. 
Specifically, reflection-first prompts exceeded their intended brevity when participants extended them into multi-step cognitive processes, and their efficacy diminished when participants required periods of undisturbed concentration. 
Prompts in this category should include an adaptive mechanism for assessing momentary state, supporting design patterns characteristic of EMIs and JITAIs~\cite{Heron2010, Nahum2017}.
However, this mechanism need not rely on sensor-based inputs. 
At a minimum, it should enable users to defer engagement or present tiered levels of complexity contingent on self-reported capacity.

\subsection{Rewriting as Another Approach to Personalization}
Content personalization can be achieved in several ways.
EMIs and JITAIs leverage sensing, models, and decision rules to identify the optimal moments when someone should receive content~\cite{Nahum2017, Mishra2021}. 
Bandit algorithms with and without sensing components have also been successfully applied to tailor intervention content to individuals' preferences \cite{kumar2024using, rabbi2019optimizing}.
Moreover, recent work has discussed the possibility of personalizing content based on users' typed responses to messages~\cite{balaskas2024designing}.

We initially asked participants to rewrite prompts as a way to provide feedback on their content, but we quickly found that this task offered its own benefits.
Participants used rewriting to shape tone, guidance, and emotional fit, effectively turning generic prompts into personal scripts aligned with their motivations and daily rhythms. 
This presents an opportunity to shift personalization from system inference to user co-authorship, thereby enabling users to explicitly articulate their preferences.

Generative AI can play a role in this regard, as large language models could help users rewrite prompts while preserving their core design principles.
This could be implemented as a lightweight editing step after prompt delivery. 
For example, users could be allowed to quickly revise the wording, swap suggested actions for more feasible alternatives, or save a preferred message for later reuse. 
Over time, the system would not need to infer a complete user model, but could instead reuse these user-authored versions as a small personal library of prompts that better fit the user's tone, routines, and constraints.
\revision{However, the generated output would need to be restricted for safety concerns to avoid inappropriate suggestions, especially for prompts related to strenuous activities, nutrition, or other health-altering actions (e.g., suggesting high-glycemic foods to someone with diabetes). Future AI-based systems should therefore have guardrails in place to enforce adherence to established medical guidelines.
}

\subsection{Limitations \& Future Work}
Our study design has some limitations worth noting.
Participants received prompts over two weeks, so we cannot extrapolate whether one-minute interventions can sustain engagement beyond that time.
Our short study duration was intentional to isolate acceptability and design features, and we observed that engagement remained fairly consistent throughout the study.
Nevertheless, efficacy should be evaluated in a longer-term study that assesses domain-appropriate outcomes during the intervention period, engagement levels over extended periods, and potential lasting benefits once prompting ceases.

Different study designs also may have yielded complementary data on engagement.
We examined video click-through rates and self-reported completion to identify broad patterns of responsiveness across conditions rather than fine-grained psychological shifts. 
We did not directly measure psychological mediators such as self-efficacy or intrinsic motivation. 
We also did not implement a mechanism to verify whether participants completed the suggested actions, as it would have required additional effort such as wearing a smartwatch to track physical activity or completing mood surveys.
Nevertheless, such data may be useful for explaining and confirming why different prompts felt more meaningful for different users.

Finally, we recognize that our participant group was small and homogeneous (i.e., North American and already comfortable with messaging apps), which limits the generalizability of our findings across cultures and levels of digital literacy.
Our findings on user rewrites and personalization suggest fertile ground for generative AI systems that co-adapt prompts with users \cite{Haag2025}, so incorporating diverse voices is crucial for ensuring that these adaptation mechanisms are robust and equitable across various user demographics and digital proficiencies.

\section{Conclusion}
In this work, we investigated one-minute interventions: messages that prompt an action that can be completed in under one minute without access to user context or sensing.
Motivated by evidence that entry and onboarding friction can hamper engagement across behavior change domains, we explored whether prompts with high trialability delivered through an everyday messaging platform could remain actionable even when timing and content were not personalized.
Keys to their success in our study were clear endpoints, reward-focused framing, and optional scaffolding.
More importantly, our work identified four types of frictions—temporal, physical, resource, and cognitive—that impeded completion, and these frictions appeared in different frequencies across domains. 
While these challenges can be addressed with heavy instrumentation and sensing, we found that allowing participants to rewrite prompts turned generic messages into personal scripts that better matched their moment and motivation.
These findings contribute empirical evidence and design suggestions for one-minute interventions as a lightweight, scalable unit of interaction—highlighting when one-minute prompts succeed, where they break down, and how designers can balance low friction with domain-appropriate adaptation.

\begin{acks}
This work was partially supported by grants to Joseph Jay Williams: NSF grant 2209819, Office of Naval Research grant number \#N00014-21-1-2576, and Natural Sciences and Engineering Research Council of Canada (NSERC) grant \#RGPIN-2019-06968. 

\end{acks}

\bibliographystyle{ACM-Reference-Format}
\bibliography{healthy}

\appendix
\onecolumn
\newpage
\label{appendix:study2-prompts}

\begin{table*}[h]
  \caption{One-minute interventions by category and flow.}
  \label{tab:study2}
  \centering 
  \begin{tabular}{p{0.08\textwidth}p{0.1\textwidth}p{0.7\textwidth}}
\toprule
\textbf{Habit Category} & \textbf{Intervention Flow} & \textbf{Intervention Example} \\
\midrule
Physical Activity & Immediate Action &
\texttt{Roll your shoulders, stretch your neck, and release tension—right now! Movement relieves stress and refreshes your focus. Follow this quick 60-second stretch routine! [Video Link].} \\

Physical Activity & Immediate Action &
\texttt{Run or walk up a flight of stairs or in place—feel your heart pump! Just 60 seconds of movement can boost circulation and refocus your mind. Hit play and go! [Video Link].} \\

Physical Activity & Reflection-First &
\texttt{How does your body feel right now? Are you holding tension anywhere—shoulders, back, or legs? Take a moment to notice, then follow this stretch to release it. Movement starts with awareness! [Video Link].} \\

Physical Activity & Reflection-First &
\texttt{What’s your posture like right now? Are you slouching, tensed up, or sitting comfortably? Take a moment to realign with this quick posture reset—small changes make a big difference! [Video Link].} \\
\midrule
Mental Well-Being & Immediate Action &
\texttt{Take 1 minute to sit still, close your eyes, and breathe deeply. Watch this short video to guide you: [Video Link]. Research shows that even one minute in meditation is a simple remedy to unease and to restore your calm.} \\

Mental Well-Being & Immediate Action &
\texttt{Click on this less than 1 minute video to hear questions that guide you through gratitude: [Video Link] Take 1 minute to pause and reflect on the past week. What’s one thing, big or small, that you are grateful for? Write it down or simply think about it for 1 minute. Gratitude will increase your happiness level.}\\

Mental Well-Being & Reflection-First &
\texttt{What is one challenge you’re facing right now? Take a moment to think about it. Now, imagine the best possible outcome—what does that look like? Watch this short video to guide you through optimistic thinking and discover a small step you can take today: [Video Link].} \\

Mental Well-Being & Reflection-First &
\texttt{What emotion is sitting with you right now? Take a deep breath and notice—are you feeling calm, stressed, or something in between? Watch this 60-second grounding video to help bring clarity to your emotions. Small check-ins like this strengthen emotional awareness and reduce mental fatigue.[Video Link].} \\
\midrule
Eating Healthy & Immediate Action &
\texttt{Swap sugary drinks for water or tea. Need help choosing a better alternative? Click here for quick sugar-free swaps [Video Link]. Cutting added sugar helps balance energy and avoid crashes!} \\

Eating Healthy & Immediate Action &
\texttt{Pick one healthy add-on to your next meal (a veggie, a protein, or a fiber-rich food). Need inspiration? Click here [Video Link] for quick meal upgrades.
A small nutrition boost helps fuel your body and mind better! } \\

Eating Healthy & Reflection-First &
\texttt{Can’t remember the last time you had water? Your body may need it more than you think. Go ahead, grab a glass or take a few sips right now and notice how it feels. Give it a try—you might find it’s exactly what you needed! Click on this link to guide you through it [Video Link].} \\

Eating Healthy & Reflection-First &
\texttt{Take a minute to think about the last thing you ate—how did it make you feel? Now, take a quick look in your kitchen, fridge, or even your bag if you’re out and about. What’s one healthy snack you could grab the next time you’re hungry? Reflect on how it could give you more energy and make you feel good. Watch this short video to guide you through mindful snacking and making small, healthy choices: [Video Link].} \\
\bottomrule
\end{tabular}
\end{table*}
\newpage
\label{appendix:flow}
\begin{figure}[h]
    \centering
    \includegraphics[width=0.45\textwidth, height=0.9 \textheight]{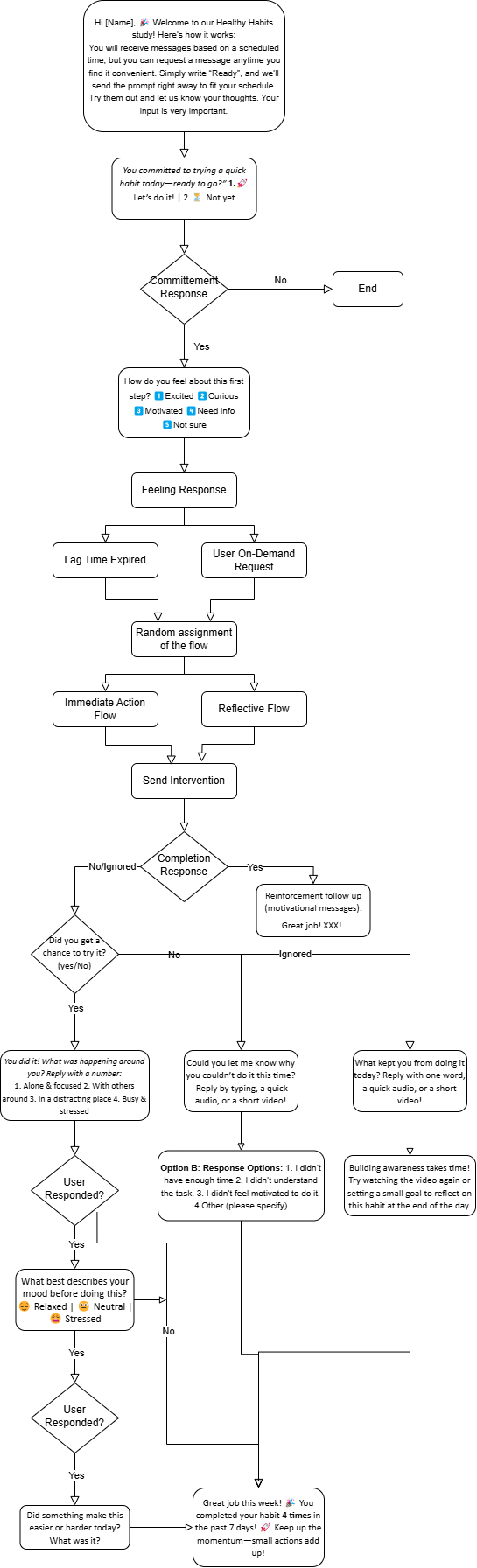}
   \caption{The steps followed by the WhatsApp bot to deliver one-minute intervention messages.}
    \Description{A flowchart showing participant interactions in the WhatsApp study, including initial onboarding, randomized intervention flows, engagement tracking, and follow-ups.}
    \label{fig:flowchart}
\end{figure}
\twocolumn
\end{document}